\documentclass[12pt]{iopart}

\usepackage{iopams}\usepackage{graphicx}  
\begin{document}

\newcommand{\kt}[1]{\ensuremath{|#1\rangle}}
\renewcommand{\br}[1]{\ensuremath {\langle #1|}}
\newcommand{\HS}{\mathcal{H}}
\renewcommand{\Tr}{\mathrm{Tr}}

\title[Entanglement Mechanisms in Scattering]{Entanglement Mechanisms in One-Dimensional Potential Scattering}

\author{N L Harshman and P Singh}

\address{Department of Computer Science, Audio Technology, and Physics, American University, 4400 Massachusetts Ave NW, Washington, DC, 20016 USA}
\ead{harshman@american.edu}
\begin{abstract}
When two non-relativistic particles scatter in one dimension, they can become entangled.  This entanglement process is constrained by the symmetries of the scattering system and the boundary conditions on the incoming state.  Applying these constraints, three different mechanisms of entanglement can be identified: the superposition of reflected and transmitted modes, momentum correlations of the reflected mode due to inversion of the relative momentum, and momentum correlations in the transmitted and reflected modes due to dependence of the scattering amplitude on the relative momentum.  We consider three standard potentials, the hard core, Dirac delta, and double Dirac delta, and show that the relative importance of these mechanisms depends on the interaction and on the properties of the incoming wave function.  We find that even when the momenta distributions of the incoming articles are sharply peaked, entanglement due to the momentum correlations generated by reflection can be quite large for particles with unequal mass.
\end{abstract}

\maketitle

\section{Introduction}

Before two particles scatter, they are in uncorrelated states.  If we assume each particle can be described by a pure state $\kt{\phi_i}$  when the particles are far apart before the interaction, then the initial total state of the system is the product of the two one-particle states $\kt{\phi_{in}}=\kt{\phi_1}\otimes \kt{\phi_2}$.  As the particles approach, the state evolves and interparticle separability is lost.  Separability does not return even if the interaction is elastic and after a long time the particles are far removed beyond the interaction region.  The boundary conditions of scattering are inherently time-asymmetric from the perspective of entanglement: practically, entangled particles cannot be easily prepared, and even if they could be, they would not approach, interact, and then emerge in a separable state.
The amount of entanglement generated by scattering is also constrained by symmetry.  For non-relativistic scattering, free particles are associated to projective representations of the Galilean group extended by mass.  Galilean invariant interactions imply conservation principles, and these shape the mechanisms by which the entanglement occurs.

For simplicity, we consider the generation of entanglement in the non-relativistic scattering of two structureless, distinguishable particles in one dimension.  Within this limited context, we show that the constraints of symmetry and time-asymmetric boundary conditions on the the incoming state imply that scattering entanglement proceeds by the combination of three mechanisms.  The simplest is the entanglement between transmission and reflection.  This mechanism of entanglement is also the coarsest, because as long as the transmitted and reflected modes are orthogonal, each particle can effectively be thought of as a two-level system, with each level corresponding to a different side of the interaction region.  The other two mechanisms create entanglement by distorting the wave functions of the modes in a non-separable fashion.  If particles have different masses, the wave function of the the reflected mode is distorted by the transformation that reverses the direction of the relative momentum.  In other words, because of the conservation of total momentum, momentum correlations between the particles are created by reflection since their momentum distributions have a finite width.    Finally, the scattering amplitudes typically vary with the relative momentum and this causes momentum correlations.  The wave function of both the transmitted and reflected mode in the outgoing state can be inseparably distorted by this effect.   The contribution of this mechanism to the entanglement of the outgoing state diminishes as each particle's wave function becomes sharply peaked about a central value, but we will show that the entanglement due to the reflection mechanism does not.

The study of how entanglement is generated in scattering has interest for a variety of reasons.  Some leading possibilities for practical implementations of quantum information processes, such as ultracold atoms and some solid state devices, are physical systems where scattering is central to the dynamics.  Also, quantum information theory with continuous variables and mixed continuous-discrete variables has many open questions, and scattering systems provide a rich structure for exploration of such systems.  Finally, scattering is a fundamental method of interaction for systems at all quantum scales, and one could hope that entanglement might provide new perspective on this basic interaction process.

Entanglement generation in non-relativistic scattering of structureless, distinguishable particles has been considered previously~\cite{schulman98,mack02,schulman04,law04,tal05,wang05, wang06, janzig06,frey07}.  Most of these previous treatments consider interactions of a single species~\cite{mack02,law04,tal05,wang05,wang06,frey07}, and have focused on particular special cases.  This article provides a unified treatment of all these results and provides a framework for further exploration and generalization.  The general methods applied here were developed in \cite{harshhutt}, in which the effect of linear transformations of observables on the entanglement of a wave function is described.  In \cite{harshhutt}, this technique is used to prove that entanglement with respect to certain sets of observables (such as internal-external entanglement, but not interparticle entanglement) is conserved in scattering, and this is applied to the mechanism of entanglement due to reflection.

In section 2, this article will show that by combining the results of \cite{harshhutt} with the constraints of symmetry and the time-asymmetric boundary conditions of scattering, the three mechanisms for entanglement can be identified.  In section 3, the interplay of these mechanisms is explored for three finite-range potentials: hard core, Dirac delta, and double Dirac delta.  Two important features emerge that contradict some assumptions found in the literature.  First, we will show that when reflection dominates the scattering interaction, even for very narrow momentum distributions, the mechanism of reflection distortion can dramatically increase the entanglement.  All that matters is the relative masses and momentum variances of the particles in the incoming state, not the overall scale, and this effect dominates at low energy.  Second, we find that although scattering resonances cause rapid variations in the scattering amplitudes and therefore the third mechanism is relevant, they do not always increase the total entanglement within the transmitted and reflected modes.

\section{Entanglement, symmetry, and scattering boundary conditions}

For two structureless, distinguishable, non-relativistic particles in one-dimension, the momentum operators of each particle $\{\hat{P}_1,\hat{P}_2\}$ form a complete set of commuting observables (CSCO).  A tensor product structure corresponding to this CSCO is $\HS_1\otimes\HS_2$, where $\HS_i$ is the single, free-particle Hilbert space.  A pure state of the system is described by the bi-momentum wave function $\phi(p_1,p_2)$.  For simplicity, we shall restrict considerations so that $\phi(p_1,p_2)\in\mathcal{S}(\mathbb{R}^2)$, the Schwartz space on the bi-momentum plane $(p_1,p_2)$.  This means that we consider wave functions that are smooth, infinitely differentiable, and rapidly decreasing at infinity.  With this mild and physically reasonable restriction, we can employ Reimann integrals and be assured of their finiteness.

The interparticle entanglement of a general pure state $\kt{\phi}\in\HS_1\otimes\HS_2$ can be calculated by the purity of the one-particle reduced density matrix
\begin{equation}
p_{12}(\phi) = \Tr_{1}\rho^2_1 = \Tr_{2}\rho^2_2,
\end{equation}
where $\rho_i$ is the one-particle reduced density matrix
\begin{equation}
\rho_1 = \Tr_{2}(\br{\phi}\kt{\phi})
\end{equation}
For the continuous-variable wave function $\phi(p_1,p_2)$, the purity of the reduced density matrix becomes  
\begin{equation}\label{eq:mompur}
p_{12}(\phi) = \int dp_1dp_2dp'_1dp'_2 \phi(p_1, p_2)\phi^*(p'_1, p_2)\phi(p'_1, p'_2)\phi^*(p_1, p'_2).
\end{equation}
The interparticle purity $p_{12}$ is an entanglement monotone (more purity always mean less entanglement) and takes values in the interval $(0,1]$ as long as the wave function $\phi(p_1,p_2)$ is normalized.  For continuous variable entanglement, the purity is useful because, unlike the entropy of entanglement, one does not need to diagonalize the reduced density matrix to calculate $p_{12}$ numerically.  Additionally, the simple form (\ref{eq:mompur}) allows for analytic results in certain cases (see below).

Without loss of generality, all calculations can be performed in the center-of-mass (COM) reference frame where the expectation value  of the total momentum operator $\hat{P}=\hat{P}_1\otimes\hat{\mathbb{I}}_2 + \hat{\mathbb{I}}_1\otimes\hat{P}_2$ in the state $\phi_{in}(p_1,p_2)$ is zero. Galilean transformations, including global boosts and translations, are represented by unitary transformations that are local with respect to the interparticle tensor product structure $\HS_1\otimes\HS_2$ and therefore do not affect the value of the interparticle entanglement~\cite{harshosid}.  In other words, the operator that performs the boost to the COM frame factors as $U(-\langle \hat{P} \rangle)=U_1(-\langle \hat{P} \rangle)\otimes U_2(-\langle \hat{P} \rangle)$.  The value $\langle \hat{P} \rangle$ is invariant under the dynamics as long as the interaction is Galilean invariant.   

One boundary condition of a scattering experiment is that the in-state $\phi_{in}$ (formally the state in the limit $t\rightarrow -\infty$) is separable with respect to the interparticle tensor product structure.  Since $\phi_{in}(p_1,p_2)=\phi_{in,1}(p_1)\phi_{in,2}(p_2)$, one calculates that $p_{12}(\phi_{in})=1$ for every scattering system.  The dynamics will generally not preserve this separability.  For example, when two particle with spin scatter non-relativistically, one can show that even in the simplified case of central interactions and narrow momentum distributions, the set of S-matrices acting on the spin degrees of freedom that lead to separable out-states depends on the specific in-state and it is a set of lower dimension on the manifold of all possible symmetry-preserving and unitary S-matrices~\cite{harsh_pra06}.

Another boundary condition for scattering is that the particle wave functions represent states that are ``incoming'' before the scattering.
If the interaction potential has finite range, ``incoming'' suggests that the single-particle position expectation values in the in-state are on opposite sides of the potential region.  Assuming the COM reference frame, the single-particle momentum expectation values are equal in magnitude and directed toward the potential region.  Further, before the interaction the position wave function should have no (or essentially no) support in the potential region  and the momentum wave functions have support only (or essentially only) on the positive semi-axis for one particle and on the negative semi-axis for the other.
For our calculations, we consider the product of two Gaussian wave packets:
\begin{equation}\label{gaussmom}
\phi_{in}^G(p_1,p_2) =  N_1N_2 e^{ip_1 a_1}e^{-\frac{(p_1 - k)^2}{4\sigma^2_1}} e^{ip_2 a_2}e^{-\frac{(p_2 + k)^2}{4\sigma^2_2}},
\end{equation}
where $N_i = (2\pi\sigma^2_i)^{-1/4}$, $k$ is the magnitude of the momentum of each particle in the COM frame, $a_i$ are the central positions, and $\sigma_i$ are the momentum uncertainties for each particle's Gaussian.  As long as $a_1=-a_2$ is large and when $k/\sigma_i\ll 1$, this wave function satisfies at least these heuristic notions of incoming.

A more refined notion of incoming boundary conditions is the Hardy space hypothesis of A.~Bohm and collaborators~\cite{bohm}.  In that formulation, further restrictions are placed on the space of allowable in-states, which are defined by the preparation apparatus, such as an accelerator.  An alternate CSCO for two particle elastic scattering is $\{\hat{P},\hat{W},\hat{\Xi}\}$ with generalized eigenvalues of the total momentum $p\in\mathbb{R}$, the internal energy $w\in\mathbb{R}^+$, and relative momentum direction $\chi=\pm$, respectively.  Then the incoming wave function $\phi_{in}(p,w,\chi)$ is a Hardy function from below in internal energy, i.e.\ it is the boundary value on the real semi-axis of a function that is analytic in the lower-half complex plane when $w$ is extended to complex values. Additionally, the wave functions are Schwartz, giving them well-behaved smoothness and convergence properties in the internal energy and the total momentum.  Conjugate requirements apply to the wave functions that represent the out-observables, which are defined by the detectors, but these do not enter the present analysis.  It is an open question as to whether the requirements of the Hardy space hypothesis are consistent with the separability constraint on the in-state described above.  However, since the Hardy-Schwartz spaces are dense in the Hilbert space, there will always be elements as close to separable as would be physically indistinguishable.

We will not consider the intricacies of the time-dependence of the scattering entanglement.  Instead, since the in-state particles are always unentangled, any entanglement in the final out-state will have been generated in the scattering event.
The out-state (formally the state in the limit $t\rightarrow +\infty$) is found by
\begin{equation}
\phi_{out}=\hat{S}\phi_{in},
\end{equation}
where $\hat{S}$ is the scattering operator. The exact form of the S-operator can be calculated for finite-range potentials by transforming  to the COM-relative momentum coordinate system
\begin{eqnarray}\label{spmtoiem}
p &=& p_1 + p_2\\
q &=& \mu_2 p_1 - \mu_1 p_2,\nonumber
\end{eqnarray}
where $\mu_i = m_i/(m_1 + m_2)$ and solving the time independent Schr\"odinger equation in the relative momentum variable $q$~\cite{merz}.  The S-matrix in the $(p,q)$-basis of the CSCO $\{\hat{P}, \hat{Q}\}$ is
\begin{equation}\label{smat}
\br{p,q}\hat{S}\kt{p',q'} = \delta(p'-p)\left(t(q)\delta(q-q') + r(q)\delta(q+q')\right)
\end{equation}
The functions $t(q)$ and $r(q)$ are the transmission and reflection amplitudes, and unitarity implies $|t(q)|^2 + |r(q)|^2=1$.  We also note in passing that the S-operator is a local operator with respect to the tensor product structure dictated by the CSCO $\{\hat{P}, \hat{Q}\}$, and so entanglement with respect to that tensor product structure is dynamically invariant~\cite{harshprl07}.  In other words, one can define a transformed wave function $\tilde{\phi}(p,q)$ and calculate
\begin{equation}\label{eq:iepur}
p_{pq}(\phi) = \int dpdpqdp'dq' \tilde{\phi}(p, q)\tilde{\phi}^*(p', q)\tilde{\phi}(p', q')\tilde{\phi}^*(p, q').
\end{equation}
and one would find that $p_{pq}(\phi_{in}) = p_{pq}(\phi_{out})$.

Using (\ref{smat}) and transforming back to the CSCO $\{\hat{P}_1,\hat{P}_2\}$, the out-state can be expressed as the sum of a transmitted and a reflected mode
\begin{equation}\label{phout}
\phi_{out}(p_1,p_2) = \phi_{tra}(p_1,p_2) + \phi_{ref}(p_1,p_2)
\end{equation}
where 
\begin{equation}
\phi_{tra}(p_1,p_2) = t(\mu_2 p_1 - \mu_1 p_2) \phi_{in}(p_1,p_2)
\end{equation}
and
\begin{equation}
\phi_{ref}(p_1,p_2) = r(\mu_2 p_1 - \mu_1 p_2) \phi_{\overline{in}}(p_1,p_2).
\end{equation}
The wave function $\phi_{\overline{in}}(p_1,p_2)$ is the in-state wave function $\phi_{in}(p_1,p_2)$ transformed by the reflection of the internal momentum $q\rightarrow -q$.  One can show that
\begin{equation}
\phi_{\overline{in}}(p_1,p_2) = \phi_{in}(\overline{p}_1,\overline{p}_2)
\end{equation}
where $(\overline{p}_1,\overline{p}_2)$ are
\begin{eqnarray}\label{trans}
\overline{p}_1 &=& (\mu_1 - \mu_2) p_1 + 2\mu_1p_2\nonumber\\
\overline{p}_2 &=& 2\mu_2 p_1 + (\mu_2 - \mu_1)p_2.
\end{eqnarray}

The wave functions $\phi_{tra}(p_1,p_2)$ and $\phi_{ref}(p_1,p_2)$ are orthogonal modes.  The domain of support for $\phi_{in}(p_1,p_2)$ (and therewith $\phi_{tra}(p_1,p_2)$) can be chosen without loss of generality as the region where $p_1 >0$ and $p_2<0$.  The domain of support of $\phi_{\overline{in}}(p_1,p_2)$ (and therewith $\phi_{ref}(p_1,p_2)$) is then the region where $(\mu_1 - \mu_2) p_1 + 2\mu_1p_2 >0$ and $2\mu_2 p_1 + (\mu_2 - \mu_1)p_2<0$.  Remembering $\mu_2 = 1 - \mu_1$ and $1 >\mu_1>0$, one can show these domains have no intersection.
Since the domains of support of the transmitted and reflected states are disjoint, the purity of the out-state is the sum of the purities of those two modes:
\begin{equation}\label{split}
p_{12}(\phi_{out}) = p_{12}(\phi_{tra})+p_{12}(\phi_{ref})
\end{equation}
From this observation, two distinct types of entangling mechanisms can be identified.  One source of entanglement is the superposition of the transmitted and reflected modes.  As long as there is not perfect reflection or perfect transmission, we will find $p_{12}(\phi_{out})<1$.
The other is the entanglement within the transmitted and reflected modes themselves, and these mechanism can be further refined into entanglement due to reflection distortion and entanglement due to the variation of $t(q)$ and $r(q)$ with $q$.

\begin{figure}
\centering
\includegraphics[width=0.25\linewidth]{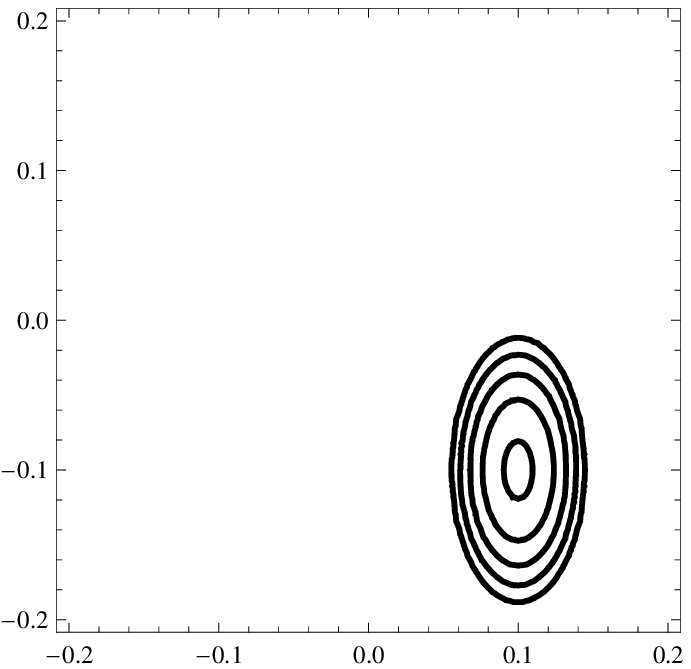}
\includegraphics[width=0.25\linewidth]{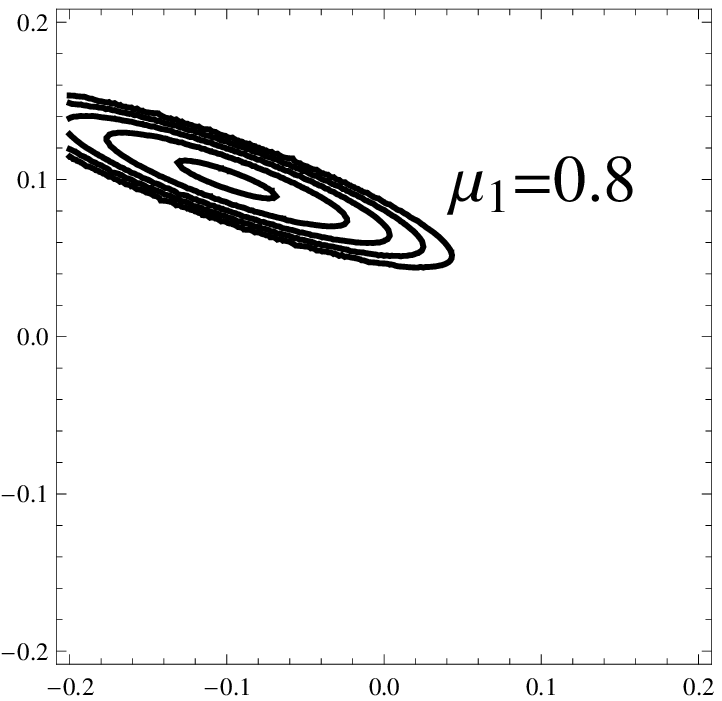}
\includegraphics[width=0.25\linewidth]{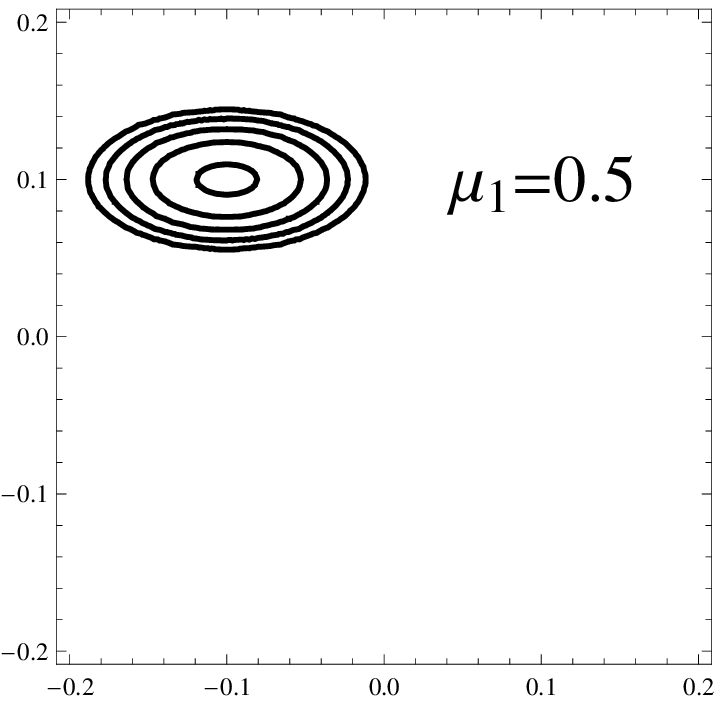}\\
\includegraphics[width=0.25\linewidth]{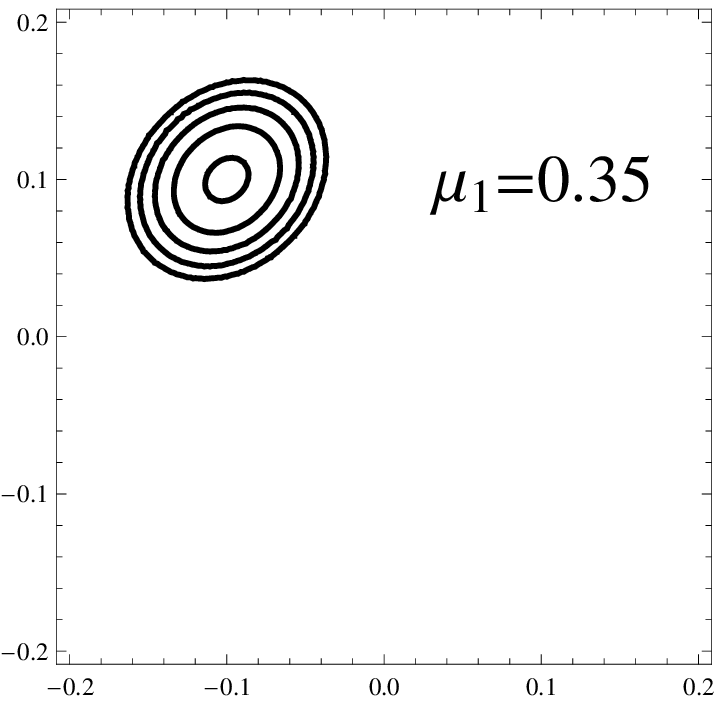}
\includegraphics[width=0.25\linewidth]{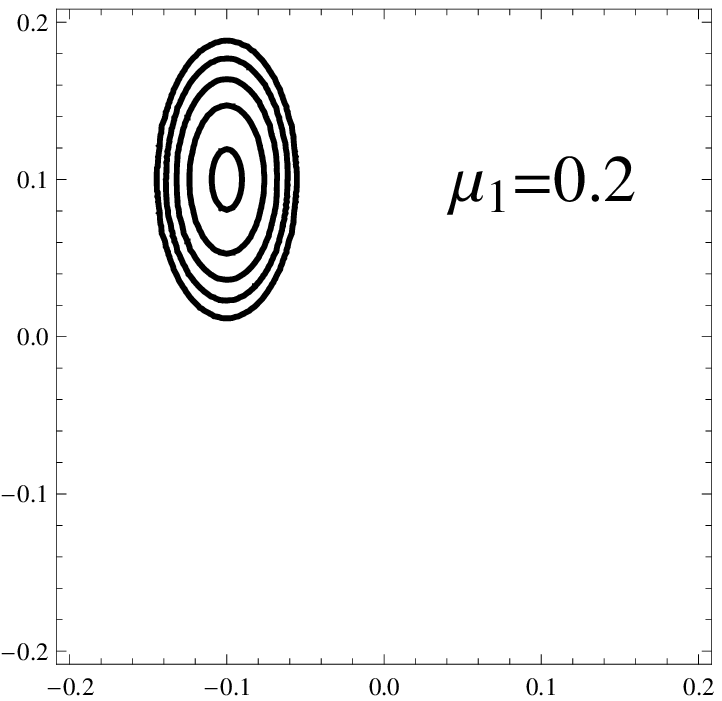}
\includegraphics[width=0.25\linewidth]{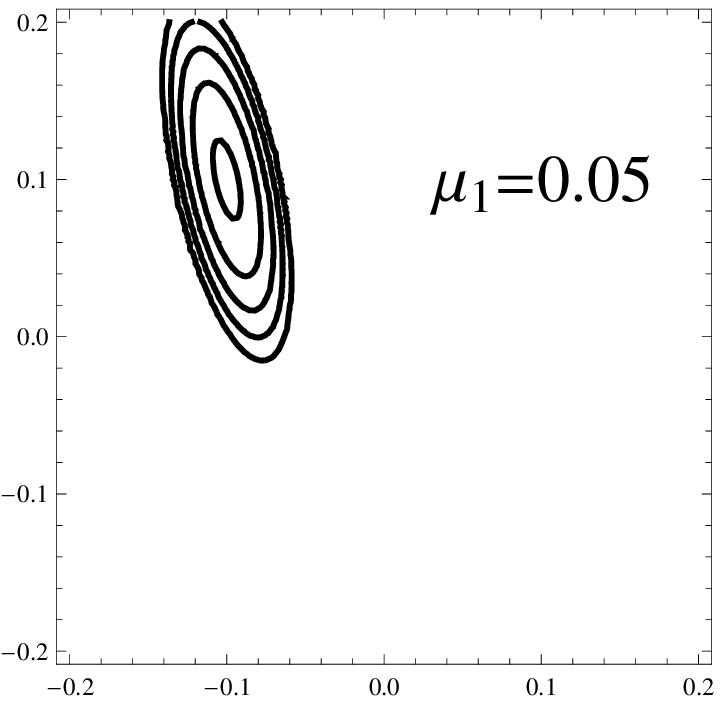}
\caption{These contour plots depict the bi-momentum probability densities $|\phi_{in}^G(p_1,p_2)|^2$ (upper, left subfigure) and  $|\phi_{\overline{in}}^G(p_1,p_2)|^2$ (the rest) for various values of the mass fraction $\mu_1=m_1/(m_1 +m_2)$ plotted on the $(p_1,p_2)$ plane.  For each graph, the central momentum is $k=0.1$ and the momentum variances are $\sigma_1 = k/10$ and $\sigma_2 = k/5$.  As the contours move outwards, each represents a reduction of probability density by a factor of 10.  Unless the major and minor axes of the ellipses align with the $(p_1,p_2)$-axes, the wave function has interparticle entanglement.}
\label{fig1} %% label for entire figure
\end{figure}

To understand the effect of reflection, note that the transformation of momentum variables (\ref{trans}) typically distorts the shape of the wave function and disrupts separability (see Figure 1).  Because a given value of relative momentum $q$ may overlap with a section in the $(p_1,p_2)$-plane, momentum correlations ensue when the transformation $q\rightarrow -q$ occurs.  However, there are some special cases when this mechanism does not create entanglement.  For the case of equal masses $\mu_1 = \mu_2 = 1/2$, we find $(\overline{p}_1,\overline{p}_2) \rightarrow (p_2,p_1)$ and so the function $\phi_{\overline{in}}(p_1,p_2) \rightarrow  \phi_{in}(p_2,p_1)$ is still separable for any wave function $\phi_{in}(p_2,p_1)$ that satisfies the incoming boundary conditions.

If we restrict ourselves to Gaussian in-state wave functions $\phi_{in}^G(p_1,p_2)$, then we find that the reflected state $\phi_{\overline{in}}^G(p_1,p_2)=\phi_{in}^G(\overline{p}_1,\overline{p}_2)$ will in addition be separable if
\begin{equation}\label{schul}
m_1/\sigma_1^2 = m_2/\sigma_2^2,
\end{equation}
a relationship first noted by Schulman~\cite{schulman98}.

\begin{figure}
\label{fig2}
\begin{center}
\includegraphics{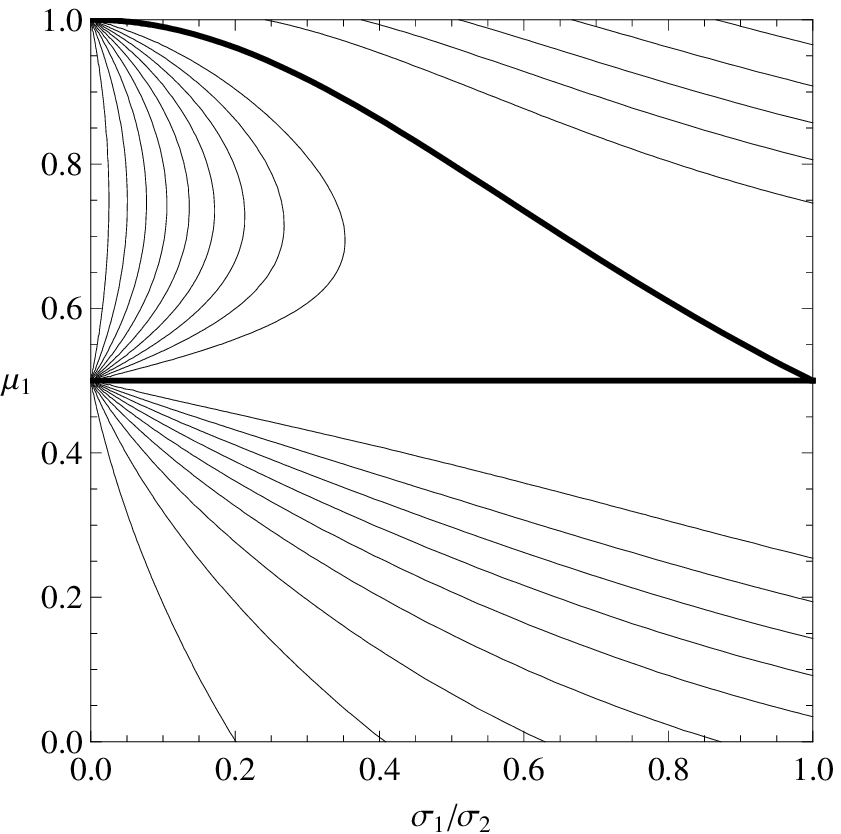}
\end{center}
\caption{This contour plot depicts the values of the function $p_{12}(\phi_{\overline{in}}^G)$ (\ref{barin2}).  The thick contours trace the two lines of maximum purity, corresponding to the values of $\mu_1$ and $c=\sigma_2/\sigma_1$ where the Gaussian bi-momentum wave function is separable even after the reflection $q\rightarrow -q$.  Each subsequent contour represents a purity reduction of $0.1$.}
\end{figure}

More generally, using the results for entanglement under linear transformations of observables found in~\cite{harshhutt}, an analytic expression for the purity $p_{12}(\phi_{\overline{in}}^G(p_1,p_2))$ can be found:
\begin{equation}\label{barin}
p_{12}(\phi_{\overline{in}}^G)=\frac{\sigma_1 \sigma_2}{\sqrt{((\mu_1 - \mu_2)^2\sigma_1^2 + 4\mu_1^2\sigma_2^2)(4\mu_2^2\sigma_1^2 + (\mu_2 - \mu_1)^2\sigma_2^2)}}.
\end{equation}
Using $\mu_2 = 1-\mu_1$ and $c= \sigma_2/\sigma_1$, this can be re-expressed as
\begin{equation}\label{barin2}
p_{12}(\phi_{\overline{in}}^G)=\frac{c}{\sqrt{((2\mu_1 - 1)^2 + 4\mu_1^2c^2)(4(1-\mu_1)^2 + (1 - 2\mu_1)^2c^2)}},
\end{equation}
and this form highlights the fact that it is the ratio of the momenta variances, and not their scale, that is important.
This function takes a maximum value of 1 when either $m_1=m_2$ or $m_1/\sigma_1^2 = m_2/\sigma_2^2$ (see Figure 2). For these particular combinations of constants, the terms in the exponential that are proportional to the product $p_1\times p_2$ all cancel out and $\phi_{\overline{in}}^G(p_1,p_2)$ is again a product of Gaussians in $p_1$ and $p_2$.  When $m_1=m_2$ the variances $(\sigma_1,\sigma_2)$ switch roles and $\phi_{\overline{in}}^G(p_1,p_2)=\phi_{in}^G(p_2,p_1)$, whereas when $m_1/\sigma_1^2 = m_2/\sigma_2^2$, we find the wave function is unchanged by reflection $\phi_{\overline{in}}^G(p_1,p_2)=\phi_{in}^G(p_1,p_2)$.

The third mechanism for entanglement is the distortion of the wave functions of the two modes due to the variation of $t(q)$ and $r(q)$ with the relative momentum $q$.  Consider the transmitted mode $\phi_{tra}(p_1,p_2)=t(q)\phi_{in}(p_1,p_2)$.  Since $q = \mu_2 p_1 - \mu_1 p_2$, the transmission amplitude $t(q)$ will generally not be a separable function of $p_1$ and $p_2$.  Each value of $q$ corresponds to a section in the $(p_1,p_2)$ plane, and these different sections will be given different weights when the in-state is convoluted with $t(q)$ to get the wave function for the transmitted mode.  Therefore, $\phi_{tra}(p_1,p_2)$ will not be separable with respect to the particle momentum variables, and momentum correlations will ensue within the transmitted mode.  A similar effect will take place in the reflected mode $\phi_{ref}(p_1,p_2)=r(q)\phi_{\overline{in}}(p_1,p_2)$ because of the inseparability of $r(q)$, except then the function $\phi_{\overline{in}}(p_1,p_2)$ may also be inseparable due to entanglement by the second mechanism described above.

\section{Results for specific potentials}

The following subsections make explicit calculations of entanglement for specific potentials.  In all cases, the in-state is assumed to have the form (\ref{gaussmom}).  This means that the in-state can be fully described by five parameters: $m_1$, $m_2$, $k$, $\sigma_1$, and $\sigma_2$.  Instead of $m_1$ and $m_2$, the following results will be expressed in terms of the mass fraction of the first particle $\mu_1 = m_1/ M$ and the total mass variable $M=m_1+m_2$.

To understand how the three different mechanisms described in the previous section contribute for these potentials in different parameter regions, we will consider the following two approximations.  The first approximation is the coarse approximation, which we will denote as approximation (C).  In (C), only the entanglement due to the superposition of transmission and reflection contributes.  The system is effectively two two-level systems, and the purity takes its minimal values (corresponding to maximal entanglement) when the uncertainty between transmission and reflection is maximal.  If $T$ and $R$ are the transmission and reflection probabilities, then 
\begin{equation}\label{approx1}
p_{12}^C(\phi_{out}) = T^2+R^2.
\end{equation}
Physically, the coarse approximation (C) will be a good approximation if two facts are true.  First, the scattering amplitudes are constants over the support of the the in-state wave function and will be evaluated at the central momentum, i.e.\ $t(q)\rightarrow t(k)$ and  $r(q)\rightarrow r(k)$.  The transmission probability is $T=|t(k)|^2$ and the reflection probability is $R=|r(k)|^2$.  Then the third mechanism does not apply since the scattering amplitudes are constants (and therefore do not disrupt separability).  Second, in the coarse approximation (C) we neglect entanglement due to reflection distortion, which is reasonable if the particles have equal masses or the if the in-state is approximately Gaussian with masses and variances satisfying (\ref{schul}).  

In the second approximation, called (C+R), we continue to assume that the scattering amplitudes are essentially constants, but we allow for entanglement due to reflection distortion.  For the transmitted mode, one calculates
\begin{equation}
p_{12}(\phi_{tra}) = |t(k)|^4
\end{equation}
and for the reflected mode one calculates
\begin{equation}
p_{12}(\phi_{ref}) = |r(k)|^4 p_{12}(\phi_{\overline{in}}).
\end{equation}
For Gaussian in-states like (\ref{gaussmom}), we find
\begin{equation}\label{approx2}
p_{12}(\phi_{out}) = |t(k)|^4+|r(k)|^4p_{12}(\phi^G_{\overline{in}})
\end{equation}
where $p_{12}(\phi^G_{\overline{in}})$ is the explicit function of $\sigma_1$, $\sigma_2$, $\mu_1$, and $\mu_2$ in (\ref{barin}) and does not depend on the exact nature of the potential (although $t(k)$ and $r(k)$ do).  Because $p_{12}(\phi_{\overline{in}})<1$, approximation (C+R) is always less than the coarsest approximation (C); reflection distortion can only increase entanglement in this approximation.

For the exact results for $p_{12}(\phi_{out})$, the four-dimensional integral (\ref{eq:mompur}) was calculated numerically.

\subsection{Hard core potential}

This is the simplest case of potential scattering.  In the relative variable $x = x_1 -x_2$, the potential has the form
\begin{equation}
V(x)=\left\{\begin{array}{ll}0 & x>0\\ \infty & x\leq 0\end{array}\right.
\end{equation}
Solving the Schr\"odinger equation in the relative momentum gives the trivial answer $r(q)=-1$ and $t(q)=0$ for all $q$. For this potential, with equal masses, the coarsest approximation (C) would suggest that a long time after hard core scattering there is no entanglement, and this result is reported in \cite{law04}.  However, as noted in \cite{schulman04,janzig06} and confirmed by these calculations, reflection can cause distortion.  In fact, for this case the approximation (C+R) is exact
\begin{equation}
p_{12}(\phi_{out}) = p_{12}(\phi_{ref})= p_{12}(\phi_{\overline{in}}).
\end{equation}
Figures 1 and 2 can therefore also be considered as depicting the entangling effects of hard core scattering.  Note that this entanglement is not dependent on any absolute scale such as the energy, momentum, or mass, but depends on the ratios of masses and variances.  This kind of entanglement due to reflection (which is a factor in other potential scattering results below) does not disappear in the narrow wave packet approximation, even though the scattering amplitudes are slowing varying or not varying at all.

This kind of entanglement due to momentum distribution correlations within a single mode can be expected to be difficult to measure compared to entanglement between transmission and reflection.  For direct measurement, one would need a device to measure the momentum distribution of the scattered particles that has at least a resolution smaller than the in-state momentum variances.  One possible scheme, developed for the study of atom-photon wave function entanglement in spontaneous emission~\cite{fedorov_pra05}, involves comparing wave function variances found in both single-particle and two-particle coincidence measurements of position.  The Fourier transformation between position and momentum wave functions is local with respect to the interparticle tensor product structure, so measuring position entanglement is equivalent to measuring momentum entanglement.  The practicality of this scheme would depend strongly on the specific nature of the system under investigation, and warrants further study.

\subsection{Dirac delta potential}

This potential has the form
\begin{equation}
V(x)=\alpha \delta(x).
\end{equation}
The reflection amplitude is 
\begin{equation}
r(q) = \frac{i}{\frac{k\hbar^2}{\alpha \mu}-i}=\frac{i}{\frac{k\hbar^2}{\alpha M \mu_1(1-\mu_1)}-i}
\end{equation}
and the transmission amplitude is $t(q)= 1 + r(q)$, where $\mu$ is the reduced mass $\mu=\mu_1\mu_2 M$. This potential has no resonances, and since the single bound state that occurs when $\alpha<0$ does not participate in the elastic scattering, we can replace $\alpha\rightarrow |\alpha|$ without affecting any scattering entanglement results.  This potential is considered in \cite{mack02,wang05,frey07} for the equal mass case.

\begin{figure}
\label{fig3}
\begin{center}
\includegraphics{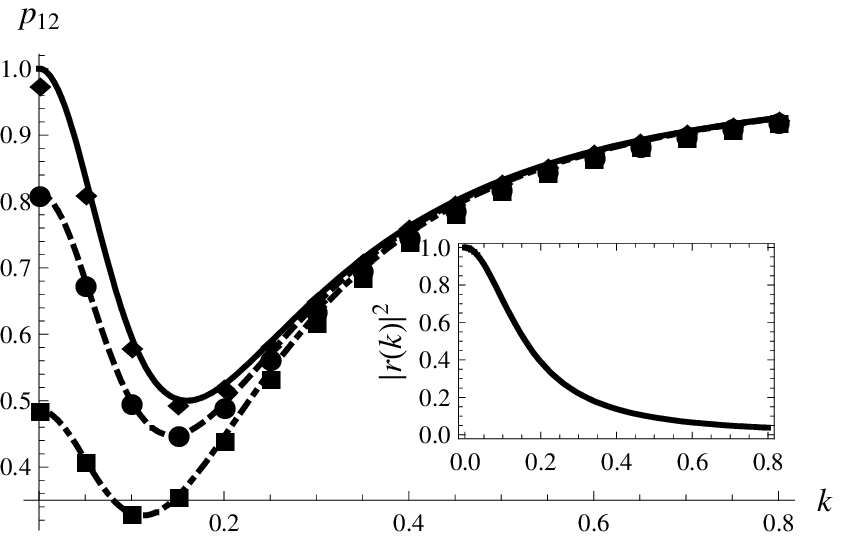}
\end{center}
\caption{The data points on the main plot depict the function $p_{12}(\phi_{out}^G)$ and the approximation (C) and (C+R) for the case $m_2= 4 m_1$ for the Dirac delta potential.  On the horizontal axis, the central momentum of the wave function $k$ is measured in units of $\alpha(m_1+m_2)/\hbar^2$.  The black line represents both the coarse approximation (C) (which is independent of $\sigma_1/\sigma_2$) and  the approximation (C+R) in the case $\sigma_1/\sigma_2=1/2$ (which satisfies the Schulman condition (\ref{schul})).  The dashed line is (C+R) for $\sigma_1/\sigma_2=1$ and the dot-dashed line is (C+R) for $\sigma_1/\sigma_2=2$.  The diamonds, circles and squares are the exact result computed numerically for $(\sigma_1=k/10,\sigma_2=k/5)$, $(\sigma_1=k/5,\sigma_2=k/5)$, and $(\sigma_1=k/5,\sigma_2=k/10)$, respectively.}
\end{figure}

To highlight the distinction between the exact result for the purity, approximation (C) and approximation (C+R), consider Figure 3 which depicts to particles scattering via the Dirac delta interaction when $m_2=4m_1$.  The figure reveals that when the particles have different masses, the relative momentum variance $\sigma_1/\sigma_2$ dramatically effects the entanglement.  Unless the condition (\ref{schul}) is fulfilled, the entanglement is enhanced by the reflection mechanism for for low $k$ because reflection dominates the scattering (see Figure 3 inset showing reflection probability $|r(k)|^2$).  Generally, the maximum entanglement (minimum purity) occurs at a value of $k$ where the transmission and reflection probabilities are equal, but the location of the extreme shifts to lower $k$ due to this reflection distortion, which for a given mass ratio, depends only $\sigma_1/\sigma_2$.  

Also note that in Figure 3, the approximation (C+R) and the exact result are very close, even for relatively wide Gaussian wave functions $\sigma_i/k = 1/5$.  The variation of $t(q)$ and $r(q)$ does not lead to much wave function distortion.  In the next example, because of resonances the scattering amplitudes vary with $k$ at a faster rate and this will no longer always be true.

\subsection{Double Dirac delta potential}

This potential has the form
\begin{equation}
V(x)=\alpha \left( \delta(x+a) + \delta(x-a)\right).
\end{equation}
The transmission amplitude is 
\begin{equation}
t(q) = \frac{q^2/b^2}{(e^{4iaq}-1)+2iq/b+q^2/b^2}
\end{equation}
and $r(q) = t(q)-1$ where $b=(m_1+m_2)\alpha/\hbar^2$.  This potential has resonant transmission $|t(q)|^2=1$ for particular values of the relative momentum $q$.  For plots below, we choose $a= 10 b^{-1}$ and measure $q$ in units of $b$.  This case is considered in \cite{tal05} for equal mass particles, and our methods clarify those results about the relative widths of resonances and the wave functions.

\begin{figure}
\label{fig4}
\begin{center}
\includegraphics{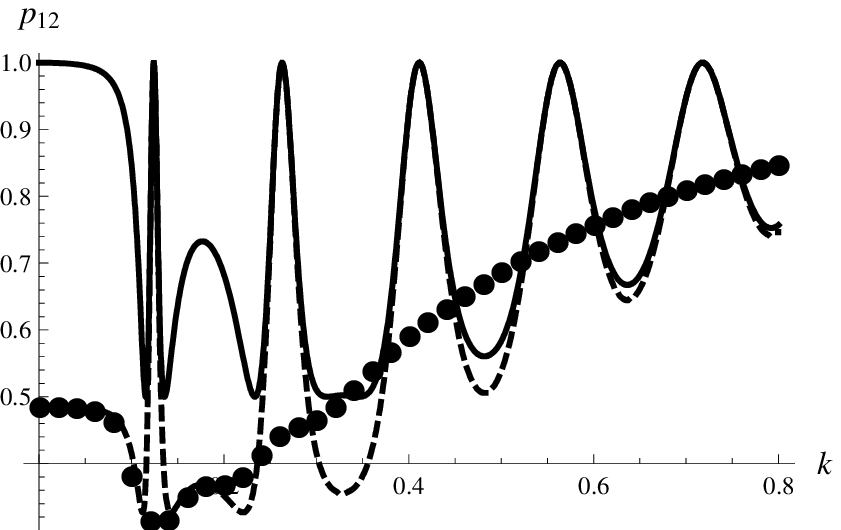}
\end{center}
\caption{The data points on the main plot depict the function $p_{12}(\phi_{out}^G)$ and approximations (C) and (C+R) for the case $m_2= 4 m_1$ for the  double Dirac delta potential.  On the horizontal axis, the central momentum of the wave function $k$ is measured in units of $\alpha(m_1+m_2)/\hbar^2$.  The black line is approximation (C) and the dashed line is approximation (C+R) for  $\sigma_1/\sigma_2=2$.  The circles are the exact result computed numerically for $\sigma_1=k/5$.}
\end{figure}

\begin{figure}
\label{fig5}
\begin{center}
\includegraphics{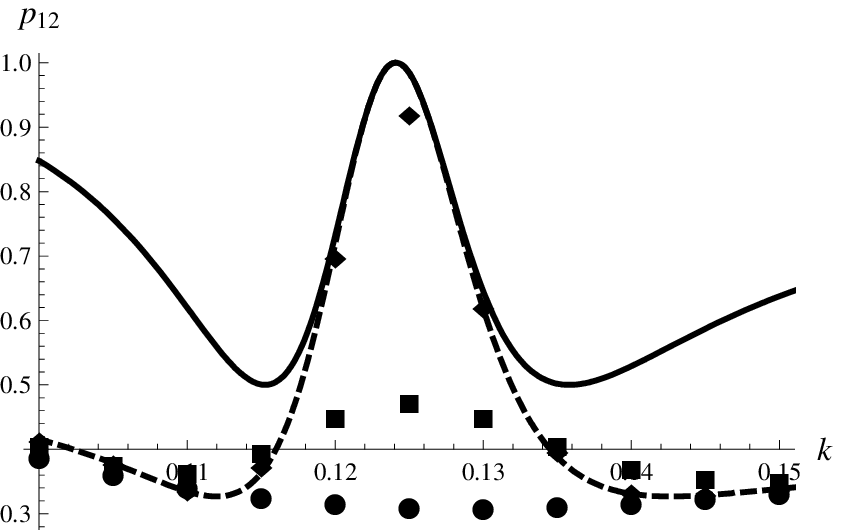}
\end{center}
\caption{This depicts an enlargement of Figure 4 around the first transmission resonance.  The circles, squares, and diamonds are the exact result computed numerically for $\sigma_1/\sigma_2=2$ with different variance scales $\sigma_1=k/5$, $\sigma_1=k/10$, and $\sigma_1=k/50$, respectively.}
\end{figure}

In Figure 4, we contrast the two approximations and the numerically-evaluated exact results for $m_2=4m_1$, $\sigma_1 = k/5$, and $\sigma_2=k/10$.  Figure 5 looks at this region around the first, narrowest resonance in more detail for a variety of absolute scales for the variances, but the same fixed ratio of variances $\sigma_1/\sigma_2=2$. At resonance, the purity of both approximations becomes unity because only the transmission mode contributes.  We see that in the exact results, the rapid variations of the out-state entanglement as found in the approximations are somewhat smoothed over.  The wider the momentum variance in the in-state, the more pronounced this effect is.  

In agreement with \cite{tal05}, we also see a slight enhancement of entanglement (reduction of purity) for wave functions wider than the narrow resonance.  As the variance scale get smaller, this effect disappears, and the exact results become closer and closer to the approximation (C+R).  Note that even as the wave functions become narrower, the coarsest approximation (C) consistently underestimates the entanglement; entanglement due to reflection is independent of the scale of the variances and only depends on the masses and the ratio of the variances.  Note that because of convolution with a fast-varying scattering amplitude, some narrow wave functions will become more entangled than wider wave functions, so the general statement `resonances increase entanglement' should be evaluated with caution.

\section{Conclusion}

In summary, by employing symmetry methods and applying the time asymmetric boundary conditions, the problem of scattering entanglement in one dimension can be analyzed by the relative importance of three different mechanisms: two-mode superposition, reflection distortion, and scattering amplitude distortion.

The overall momentum dependence of the entanglement is determined on a coarse scale by the two-mode effect, but if one could measure the momentum distributions of the two particles in the out-state, then further entanglement would be detected.  The entanglement due to reflection is intriguing because it depends on the ratio of the particle masses and the ratio of the momentum variances.  The effect is most pronounced when the more massive particle has a more certain momentum.   No matter how sharp the initial momentum distributions are, quantum correlations ensue for Gaussian states unless the Schulman condition (\ref{schul}) is satisfied or the masses are equal.  As the distribution gets narrower, however, it would also become more difficult to measure this entanglement.  Scattering amplitudes that vary rapidly with relative momentum on the scale of the variances also distort the wave function in an inseparable manner, and in contrast, this kind of entanglement becomes less prominent for narrow momentum distributions.

More complicated scattering problems must be considered if these methods will be useful for practical applications to quantum information processes with cold atoms and solid state devices.  Generally, these will require multi-dimensional results for identical particles with spin (although some one-dimensional scattering may have applications; see \cite{buscemi} for examples).  In addition to possible new effects, these three mechanisms should still be applicable to these situations.  The application of symmetry methods and the restriction of time-asymmetric boundary conditions remain valid, and they will be the starting point for future generalization of this work.

\ack

The authors would like to acknowledge the Research Corporation for supporting this research through a Cottrell College Science Award.  They also would like thank A.~Patch, S.~Wickramasekara, Y.~Strauss and I.~Antoniou for useful discussions and the organizers of 5th International Symposium of Quantum Theory and Symmetries in Valladolid Spain in July 2007 for their hard work and hospitality (especially M.~Gadella, F.~G\'omez-Cubillo, and \c{S}.~Kuru).  Finally, the comments of the anonymous referees improved this article.

\end{document}